\documentclass[%
 reprint, 
nofootinbib,
pra,
]{revtex4-2}

\usepackage{graphicx} 
\usepackage{dcolumn}
\usepackage{bm} 

\usepackage{color}
\usepackage{xcolor}

\usepackage[english
]{babel}

\usepackage{amsmath}
\usepackage[normalem]{ulem}

\begin{document}
\title{\textbf{Interplay of the channel-closing and bound-bound
transition resonances in multiphoton ionization and harmonic generation in intense laser pulses}}%
\author{
A.~D.~Krupin$^{1}$, V.~V.~Strelkov$^{2}$, and M.~Yu.~Ryabikin$^{1,3\ast}$
}
\affiliation{
\mbox{$^{1}$A.V. Gaponov-Grekhov Institute of Applied Physics of the Russian Academy of Sciences,} \\ {46 Ulyanov Street, Nizhny Novgorod 603950, Russia} \\
\mbox{$^{2}$P.N. Lebedev Physical Institute of the Russian Academy of Sciences, 53 Leninskiy Prospekt,  Moscow 119991, Russia} \\
\mbox{$^{3}$N.I. Lobachevsky State University of Nizhny Novgorod, 23 Gagarin Avenue, Nizhny Novgorod 603950, Russia} \\
$^{\ast}$m\_yu\_ryabikin@mail.ru
}
\date{\today}

\begin{abstract}
  In this paper, using a simplified model of the xenon atom, we numerically study the possibilities of efficient generation of coherent pulses in the XUV range through the resonant interaction of atoms with a moderate-intensity laser field, leading to the generation of its harmonics. We demonstrate the interplay of two systems of resonances 
  affecting the harmonic generation efficiency. One is the channel-closing resonances, which arise when the sum of ionization and ponderomotive energies is equal to the energy of an integer number of laser photons. The second is the bound-bound transition resonances corresponding to an integer number of photons with a total energy equal to the energy gap between the Stark-shifted ground and excited states. 
  The harmonic yields in the range of laser parameter values where both resonances occur exhibit a peculiar behavior, namely, near the intersection point of the resonances, a pronounced dip is observed, while the regions of increased generation efficiency due to the combined contribution of both enhancement mechanisms are slightly shifted from this point. We argue that this behavior, which is somewhat similar to the well-known phenomenon of 'avoided crossings', is associated with the formation of Fano-type resonant spectral lines. In contrast to ’avoided crossing’ phenomena known in molecular physics, in the found interplay the contribution of one resonance system can be controlled, which is useful for experiments.
\end{abstract}

\maketitle

\section*{Introduction}
Resonant effects during the interaction of an intense laser field with an atomic gas or a laser plume can significantly improve the efficiency of high harmonic generation (HHG), as demonstrated in experiments~\cite{toma1999resonance, Ganeev2006, Gilbertson_2008, Shiner2011, ackermann2012resonantly,  Rothhardt2014,  Singh2021, Singh2026}. 
Theoretical methods are well-developed  to describe an enhancement of high harmonics due to resonances with autoionizing states~\cite{Kheifets2008, Tudorovskaya_2011, Strelkov2014, Wahyutama2019, Kheifets_2020, Strelkov_2025, Romanov2024} or giant resonances~\cite{Frolov2009, Pabst, Kheifets_2019, Romanov_2021} in the case of intense generating fields corresponding to the tunneling ionization regime (the Keldysh parameter $\gamma \ll 1$). However, in many HHG experiments (including resonant HHG experiments~\cite{Singh2021}) the laser field parameters correspond to the multiphoton or intermediate ionization regimes (the Keldysh parameter $\gamma \gtrsim 1$). The theoretical description of the laser-atom interaction in this regime requires taking into account the actual structure of atomic levels. The complexity of such analytical studies makes numerical approaches especially important in this domain.

The time-dependent Schr\"odinger equation (TDSE) serves as the fundamental equation for describing HHG. Despite the modern computational capabilities, even when using single-active-electron approximation, solving the full-dimensionality (3D) TDSE remains a highly resource-intensive task. Therefore, it becomes computationally prohibitive to conduct numerical experiments that require solving the TDSE multiple times for different parameters. One way to overcome this constraint is to reduce the dimensionality of the problem. Under such conditions, preserving the correct intrinsic physical properties of the atomic system becomes essential for an adequate modeling of its interaction with the strong laser field and, in particular, HHG. This can be achieved by using low-dimensional model atomic potentials, where the essential properties of an atom are included through additional free parameters. This approach has been thoroughly explored in previous works~\cite{silaev2010strong, majorosi2020density, pedersen2025polarizability}.

In this paper, we use a 1D TDSE model to investigate HHG efficiency in a rarefied xenon gas. The low computational cost of this model allows for high-resolution scans across the laser field parameters. Consequently, we can not only identify the parameter regions leading to resonant enhancement but also examine the detailed structure of the resonances in these regions, including the case of multiple resonances and their interplay.

\section{Numerical approach} \label{1-sec}
To model a xenon atom within our reduced-dimensionality framework, we employ the modified Pöschl-Teller potential
 \begin{equation} \label{eq: potential PT}
     V_0(x) = -\frac{1}{2} \alpha^2 \lambda (\lambda - 1) \cosh^{-2}(\alpha x),
 \end{equation}
where the parameters $\alpha$ and $\lambda$ are chosen to reproduce two key physical characteristics of atomic xenon: the ground-state energy ($-12.13$~eV) and the transition energy from the ground state to the excited $5p^5(^{2}P^o_{3/2})6s$ state ($8.437$~eV). Note that this frequency is very close to the frequency relevant to the thorium-229 nuclear clock~\cite{Herrera_2013, Seiferle_2019}. So, the UV resonantly generated in xenon can be used to excite the clock transition in thorium~\cite{Tiedau}.

The analytical form of the eigenstates for the potential~(\ref{eq: potential PT}) is given in~\cite{Flugge1971}. However, a more accurate approach is to compute them numerically: this guarantees that the solutions remain stationary on the discrete computational grid, and the accuracy is then determined solely by the chosen numerical method. Here, the stationary-state problem is solved via the imaginary-time propagation method~\cite{kosloff1986direct}.

The ground-state wavefunction obtained in this manner then serves as the initial condition for the time-dependent propagation governed by the TDSE (hereinafter, the atomic units are used)
\begin{equation} \label{eq: TDSE}
    i \frac{\partial \Psi(x,t)}{\partial t} = \left[ -\frac{1}{2} \frac{\partial^2}{\partial x^2} + V_0(x) + x F(t) - iW(x)\right] \Psi(x,t),
\end{equation}
where $V_0(x)$ is the model potential defined above~\eqref{eq: potential PT}, $F(t)$ is the electric field of the laser pulse, and $-iW(x)$ is the complex absorbing potential (CAP). The equation \eqref{eq: TDSE} is solved via the split-operator method combined with the fast Fourier transform~\cite{fleck1976time}.
    In all subsequent calculations, $F(t) = E_0 \cos(\omega_0 t) \sin^2(\pi t / D)$, where $D$ is the full pulse duration; it is chosen to provide an  FWHM duration equal to $30$ field cycles for all considered frequencies.

    Unless stated otherwise, all calculations are performed inside the box $[-x_{\max}, x_{\max}]$ with $x_{\max} = 70$ and a uniform step $\Delta x = 0.1$. Such parameters satisfy all accuracy criteria, namely, those related to the resolution of the atomic potential~\eqref{eq: potential PT}, resolution of the shortest electron de Broglie wavelengths, and sufficient space to accommodate long electron trajectories.

    The CAP plays a crucial role for numerical computations because it prevents unphysical interference effects caused by wave-packet reflections from the box edges~\cite{kosloff1986absorbing}.
   The following function is chosen to implement the CAP: $W(x) = 30\cosh^{-2}\left(0.5[x_{max} - |x|]\right)$, where the amplitude and the width parameters were chosen to optimally satisfy the requirements listed above. Furthermore, the absorption induced by the CAP can be regarded as a measure of the ionization yield.

We calculate the time-dependent dipole acceleration $a(t)$ at each simulation step using the Ehrenfest theorem~\cite{burnett1992calculation}
\begin{equation} \label{eq: dipole acceleration}
    a(t) = -\left\langle \Psi(x,t) \left| \frac{\partial V_0(x)}{\partial x} + F(t) \right| \Psi(x,t) \right\rangle.
\end{equation}
Afterwards, we calculate the spectral power of the dipole acceleration $|a(\Omega)|^2$ that characterizes the HHG yield.

Note that when analyzing the harmonic spectrum, the driving force term $F(t)$ can be omitted from \eqref{eq: dipole acceleration}, as we are not interested in the response at the fundamental laser frequency. Prior to Fourier transformation, the signal $a(t)$ is multiplied by a windowing function to suppress the numerical noise arising after the laser pulse has passed, thus improving the clarity of the spectral power $|a(\Omega)|^2$. Finally, the yield of a specific harmonic is quantified by integrating $|a(\Omega)|^2$ over a narrow spectral window with a half-width of $0.15\omega_0$ centered at the corresponding frequencies $\Omega = q\omega_0$.

\section{Results} \label{2-sec}

Using the approach outlined above, we analyze the harmonic yields and ionization probability plotted as colormaps over the plane of laser fundamental frequency and peak intensity. 
To enhance visual contrast, the harmonic yields are divided by $E_0^6$. Next, we consider the high- and low-intensity cases separately to highlight the features specific to each regime.

\begin{figure} 
\raggedright
a)

    	\includegraphics[width=0.4\textwidth]{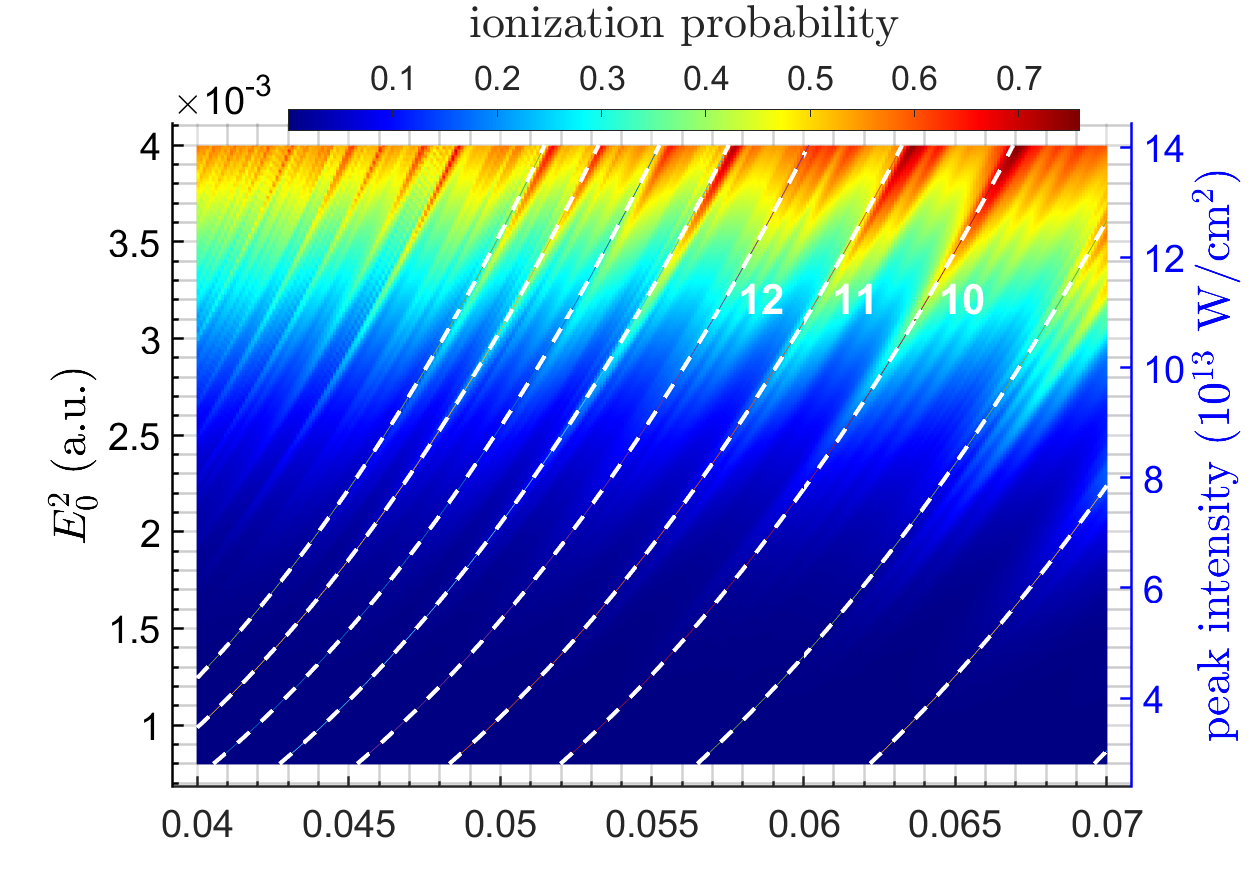}

        b)  
        
        \includegraphics[width=0.4\textwidth]{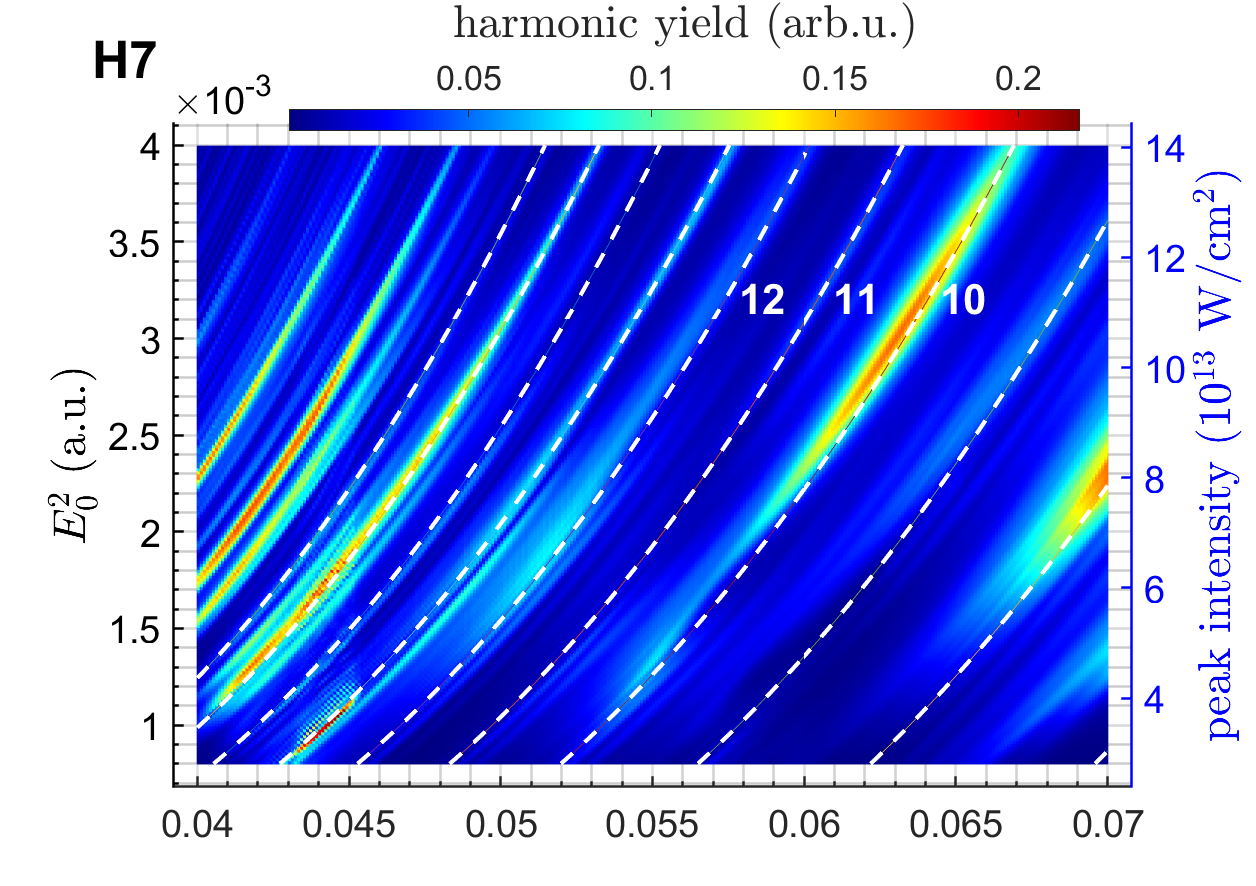}

        c) 
        
        \includegraphics[width=0.4\textwidth]{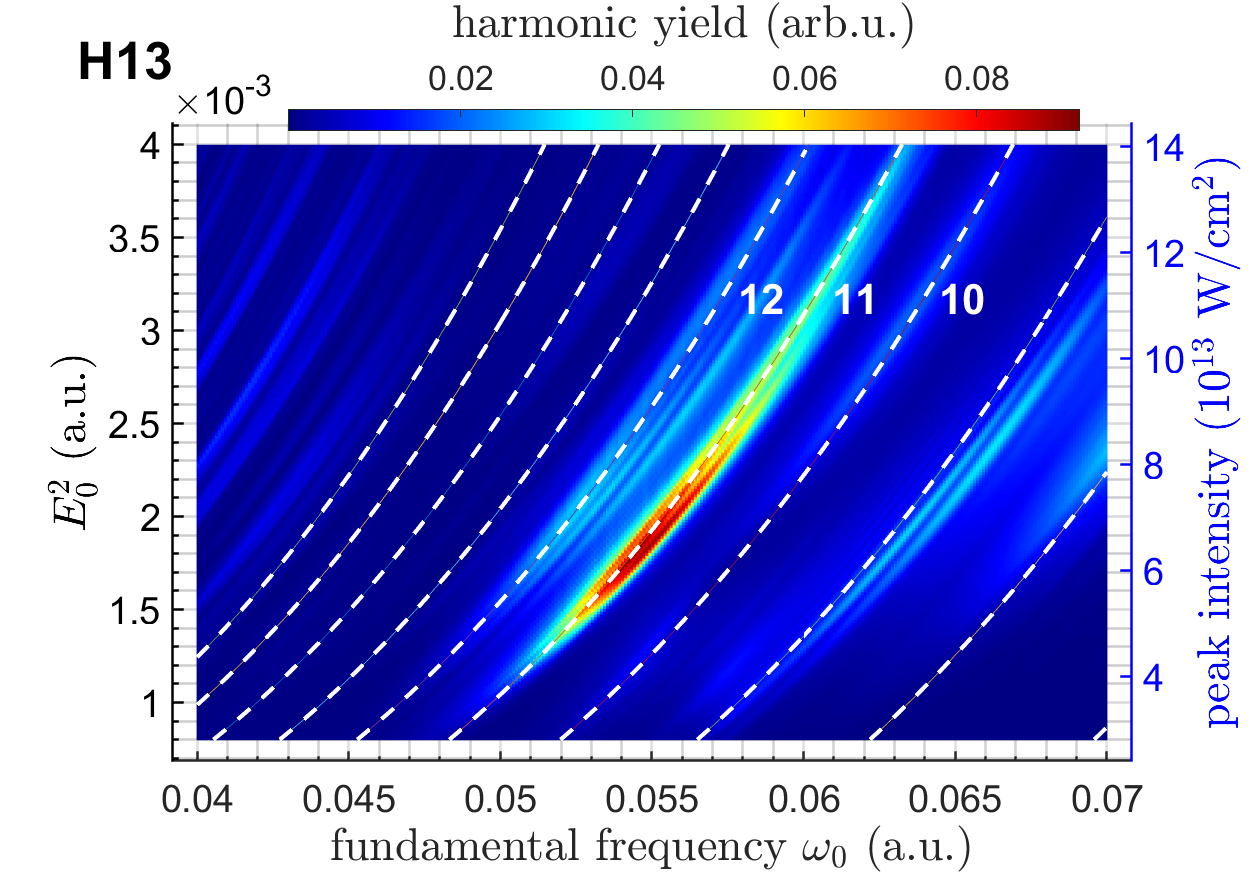}
         \caption{ 
        Ionization probability (a) and  yields of 7th (b) and 13th (c) harmonics. White dashed lines correspond to the channel-closing effect for various photon numbers $N_{cc}$ defined by Eq.~(\ref{closing}), some are labeled in the graph. The 7th harmonic is below-threshold whereas the 13th one is above-threshold for much of laser parameters shown in the graph. Harmonic yields are divided by $E_0^6$. }
	\label{fig:harmonics_and_ionization_high_intensity}
\end{figure}

\subsection{Channel closing in photoionization and harmonic generation}

We first examine the high-intensity case with peak intensities ranging from $2.8 \times 10^{13}$ to $1.4 \times 10^{14}$ W/cm$^2$. The corresponding colormaps of the ionization probability and harmonic yield (the 7th and 13th harmonics are shown as representative examples) are presented in Figure~\ref{fig:harmonics_and_ionization_high_intensity}. Both quantities exhibit a clear resonant structure. These resonances are the manifestation of the channel closing taking place when the ionization energy Stark-shifted by the ponderomotive energy  $U_p=E_0^2/4\omega_0^2$ comes into $N_{cc}$-photon resonance with the laser field: 
\begin{equation}
  I_p + U_p = N_{cc} \omega_0,  
  \label{closing}
\end{equation}
where $I_p$ is the field-free ionization energy.

Different aspects of the channel-closing phenomena are actively studied in recent decades. Early works focused on the effect of the channel closing on the ionization and photodetachment yields in the multiphoton ionization regime~\cite{Tang_1991, Becker_1990, Eberly_1991,  Delone}. Further studies investigated the influence of the channel closing on the above-threshold ionization (ATI) electron spectra~\cite{Muller_1998, Paulus_2001, Popruzhenko_2002, Milo_2007, Popov_2008}, HHG~\cite{Borca_2002, Milo_2002, Frolov_2007, Ishikawa_2009}, and the connection between channel closing and population trapping in the Rydberg states~\cite{Popov_2011, Popov_2014}. 

An increase of the overall ionization yield when condition~(\ref{closing}) is satisfied was found for both multiphoton and tunneling ionization regimes. In~\cite{Eberly_1991} this was attributed to the contribution of the first peak in the photoelectron spectrum: 'Since the lowest-energy ATD (above-threshold detachment) peak is the largest one in the spectrum, and since as it disappears at the threshold the higher-energy peaks retain relatively little probability, we expect a corresponding sharp drop in the photodetachment rate.' This explanation is definitely valid in the multiphoton and intermediate regimes, when the first peak is strongly dominating. However, in the tunneling regime, the electron spectrum consists of numerous peaks, and disappearing of just one, even the strongest one, does not necessarily affect the overall ionization yield. We propose somewhat different explanation of the deep modulation of the ionization yield. Namely, let us consider an electron wave packet that is initially in the ground state and its photoionization begins at some point in time $t_0$. After a certain number $K$ of the laser field cycles, the wave packet ends up in a continuum, having  accumulated some phase $\Phi_{pi}$. Then it moves in the continuum during one laser cycle, eventually accumulating some additional phase $\Phi_{free}$. Now let us consider another wave packet that remained in the ground state region for one laser cycle after time $t_0$, accumulating phase $- 2 \pi I_p/  \omega_0$, and was then photoionized during $K$ cycles similarly to the first wave packet, thus accumulating phase $\Phi_{pi}$ and having the same energy in the continuum. Thus, at time $t_0+2 \pi (K+1)/ \omega_0$ the two packets interfere in the continuum, their phase difference is $\Phi_{free}+ 2 \pi I_p/  \omega_0$; note that the phase $\Phi_{pi}$ cancels out. Their interference is pronounced if the first of them has not traveled far from the origin, so we can assume that the average velocity for both wave packets is zero. Then $\Phi_{free}=2 \pi U_p/ \omega_0$. The interference is constructive if the phase difference is $2 \pi N_{cc}$, thus we obtain condition (\ref{closing}).

The channel-closing fringes in Figure~\ref{fig:harmonics_and_ionization_high_intensity} are smeared out due to the realistic bell-shaped laser pulse envelope we use, but are still rather pronounced.

In Figure~\ref{fig:harmonics_and_ionization_high_intensity}, we see that the regions of enhanced harmonic generation (HG) generally coincide with ionization resonances. This behavior was reported earlier for above-threshold high-order harmonics~\cite{Borca_2002, Milo_2002, Milo_2007, Frolov_2007, Ishikawa_2009}.  The similar behavior of below-threshold harmonics, seen in Figure~\ref{fig:harmonics_and_ionization_high_intensity}, indicates that their generation is inherently linked to the electron's motion in the continuum. This can be explained by the generation of these harmonics due to frustrated tunneling~\cite{Nubbemeyer_2008, Xiong_2016} via a 'long' quantum orbit (characterized by a free-motion time $\tau_{free}$ of about one laser cycle) and a number of 'very long' orbits (with $\tau_{free}$ exceeding one cycle).


\subsection{Channel closing in above-threshold harmonic generation in the intermediate ionization regime}


Figure~\ref{fig:harmonics_and_ionization_high_intensity}(c) shows the intensity of the 13th harmonic, and similar figures for harmonics from 11th to 17th are presented in the Supplemental Material~\cite{data_Krupin}. For much of the laser frequencies and intensities shown in the figure, these harmonics are the above-threshold ones, that is, $q>N_{cc}$, where $q$ is the harmonic order and $N_{cc}$ is given by Eq.~(\ref{closing}). One can see that the HHG efficiency is very high for specific $N_{cc}$, namely, if 
\begin{equation}
   q=N_{cc}+m,
   \label{closing_2}
\end{equation}
where $m=2$ for H11-H15 and $m=4$ for H17.

When this condition is fulfilled for the 13th harmonic, it can be  more intense than the 11th one. Moreover, under this condition the 13th harmonic is more intense than when it is below-threshold or threshold one ($N_{cc} \ge 13$) as well as when it is above-threshold one for higher fundamental frequencies ($N_{cc}=10,9,8$). The latter feature is remarkable especially in the region where the harmonic is below the $I_p+3.17 U_p$ cut-off, because this feature is in contrast with very typical increase of HHG generation efficiency with the fundamental frequency in this region.

For the tunneling ionization regime, enhanced generation of the above-threshold harmonics under the channel-closing condition~(\ref{closing}) was discovered in~\cite{Milo_2007} and explained by the constructive interference of multiple quantum trajectories. Generation due to multiple trajectories ('short', 'long', and multiple 'very long') is possible for the lower-frequency part of the plateau, namely for harmonics with photon energies below $I_p+2.4 U_p$, see Fig.~1 in~\cite{Lewenstein}. Moreover, for harmonics from  $I_p+1.5 U_p$ to  $I_p+2.4 U_p$, 'very long' trajectories have caustics (similar caustics in a two-color field were studied in~\cite{Raz2012, Birulia_2019}), which can make the contribution of these trajectories more pronounced. Thus, observed maxima given by Eq.~(\ref{closing_2}) can be attributed to the constructive interference of several caustics due to  'very long' trajectories.




\begin{figure}
\begin{center}
    	\includegraphics[width=0.4\textwidth]{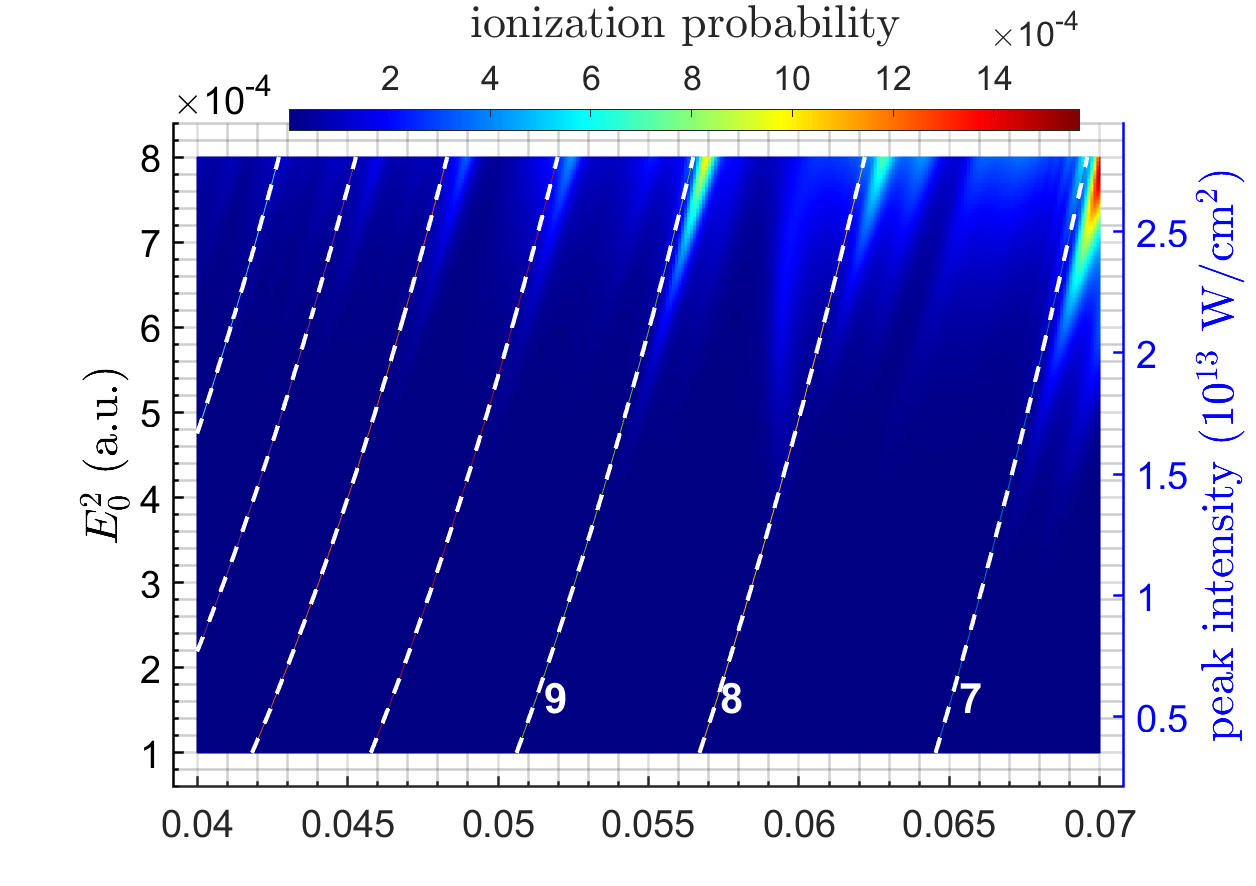}
        \includegraphics[width=0.4\textwidth]{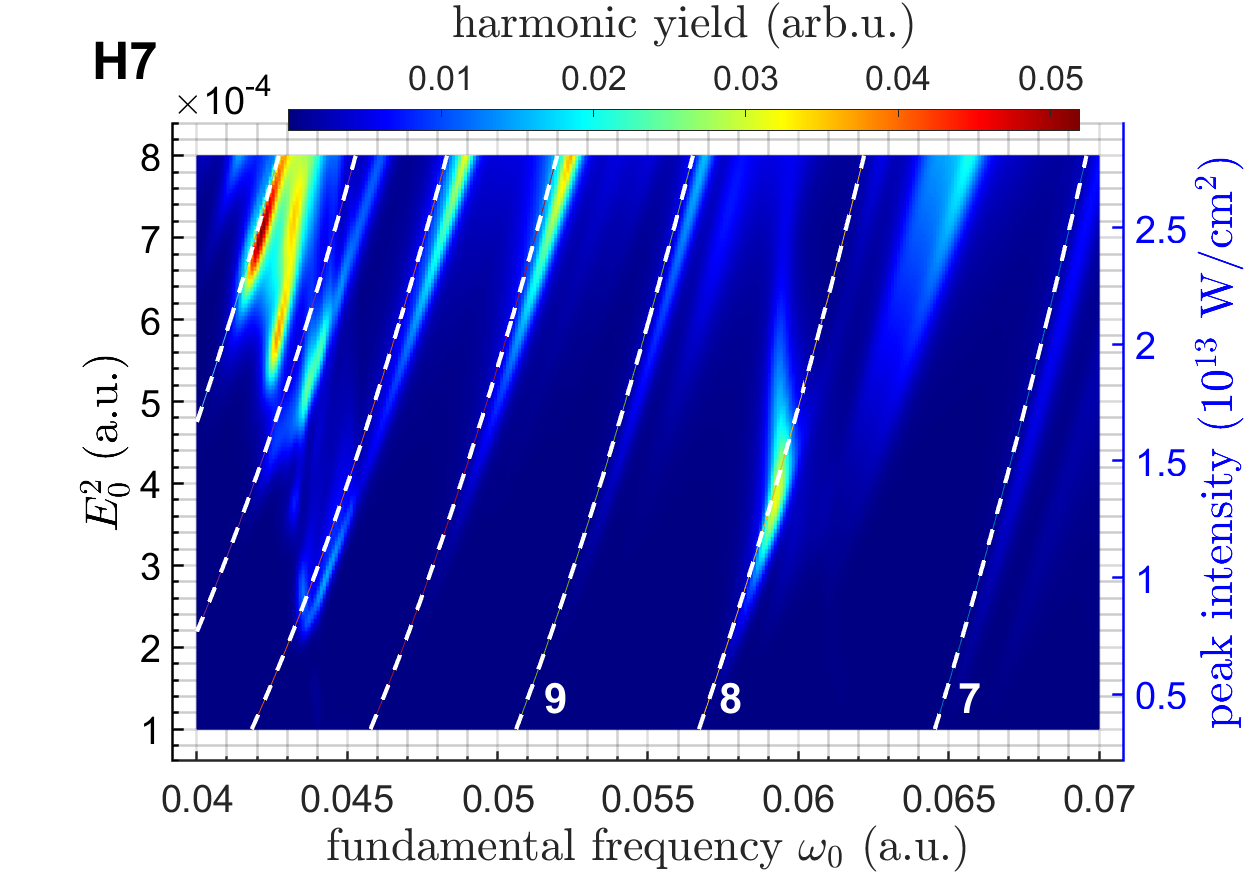}
	\caption{The same as Fig. \ref{fig:harmonics_and_ionization_high_intensity}(a,b) but for lower laser intensities.}
	\label{fig:harmonics_and_ionization_low_intensity}
\end{center} 
\end{figure}

\begin{figure*}
\begin{center}
    	\includegraphics[width=0.3\textwidth]{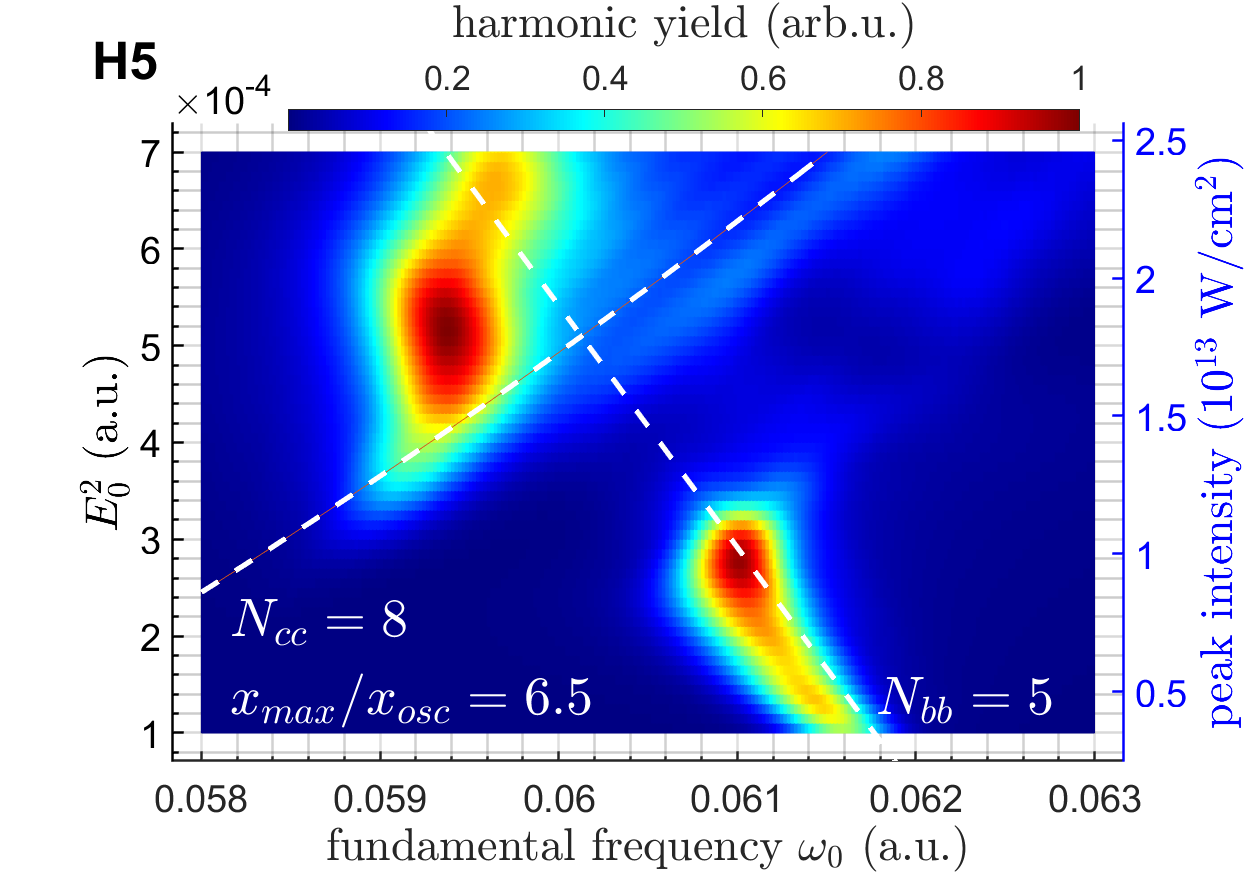}
        \includegraphics[width=0.3\textwidth]{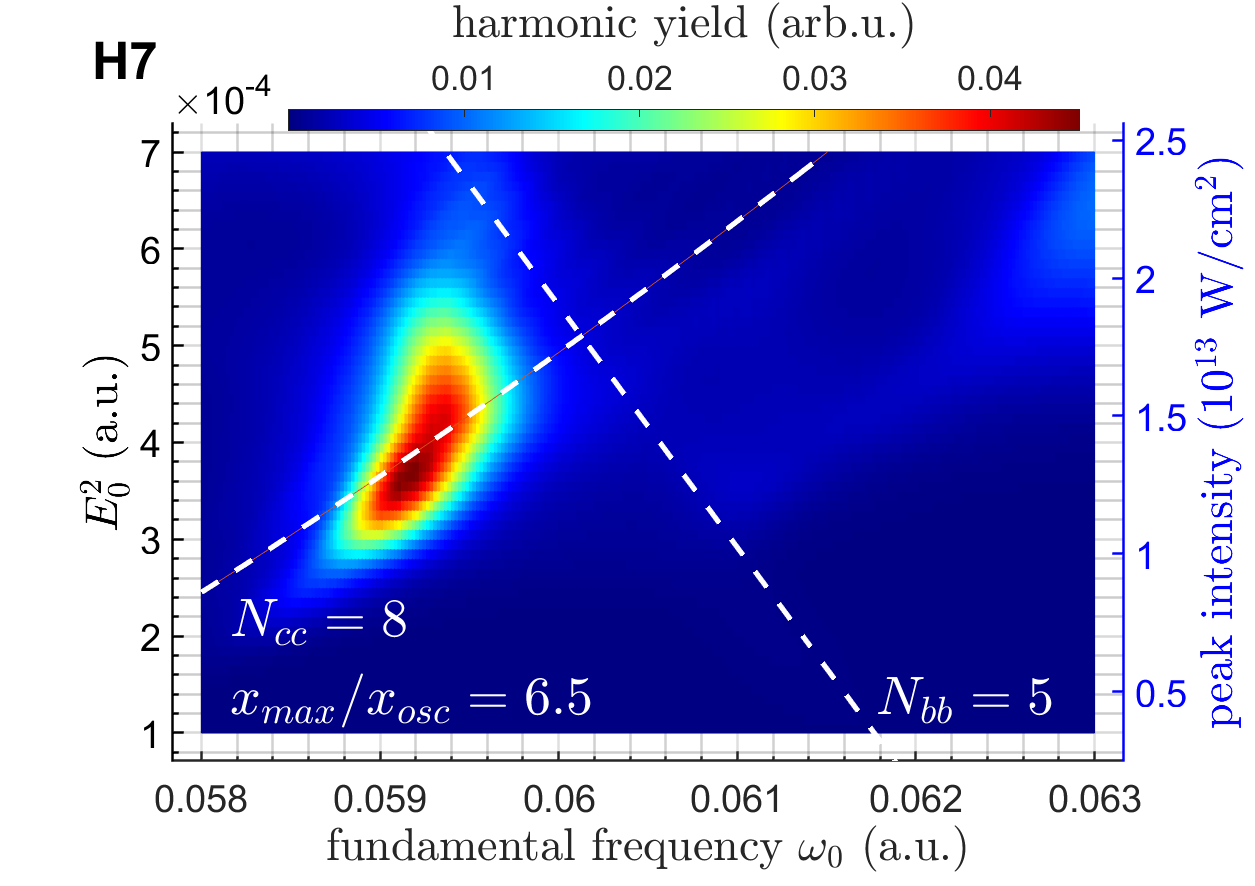}
        \includegraphics[width=0.3\textwidth]{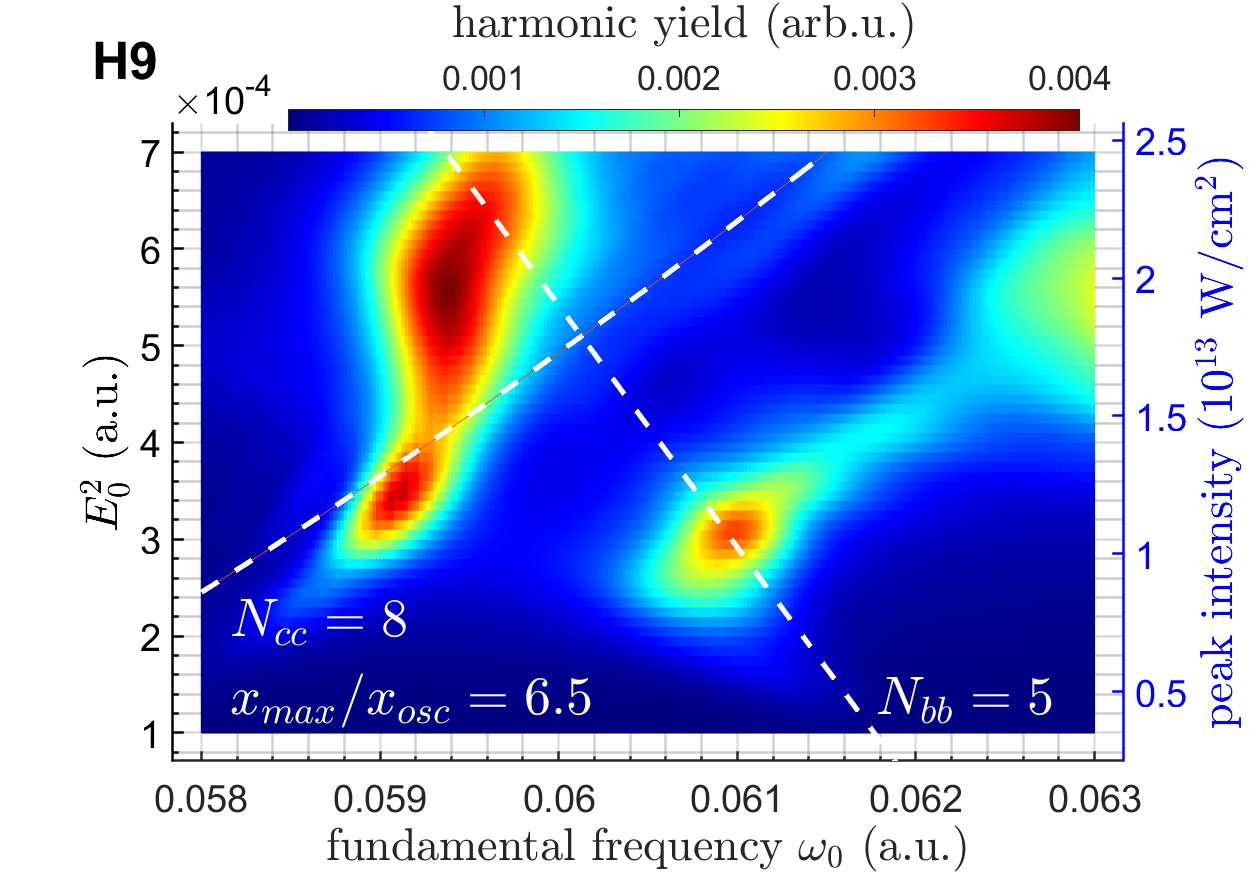}
	\caption{The same as Fig.~\ref{fig:harmonics_and_ionization_low_intensity} for H5, H7, and H9 in the region of interference between channel-closing and bound–bound transition resonances.}
	\label{fig:avoided_crossing_area}
\end{center} 
\end{figure*}

\begin{figure}
\begin{center}
	\includegraphics[width=0.4\textwidth]{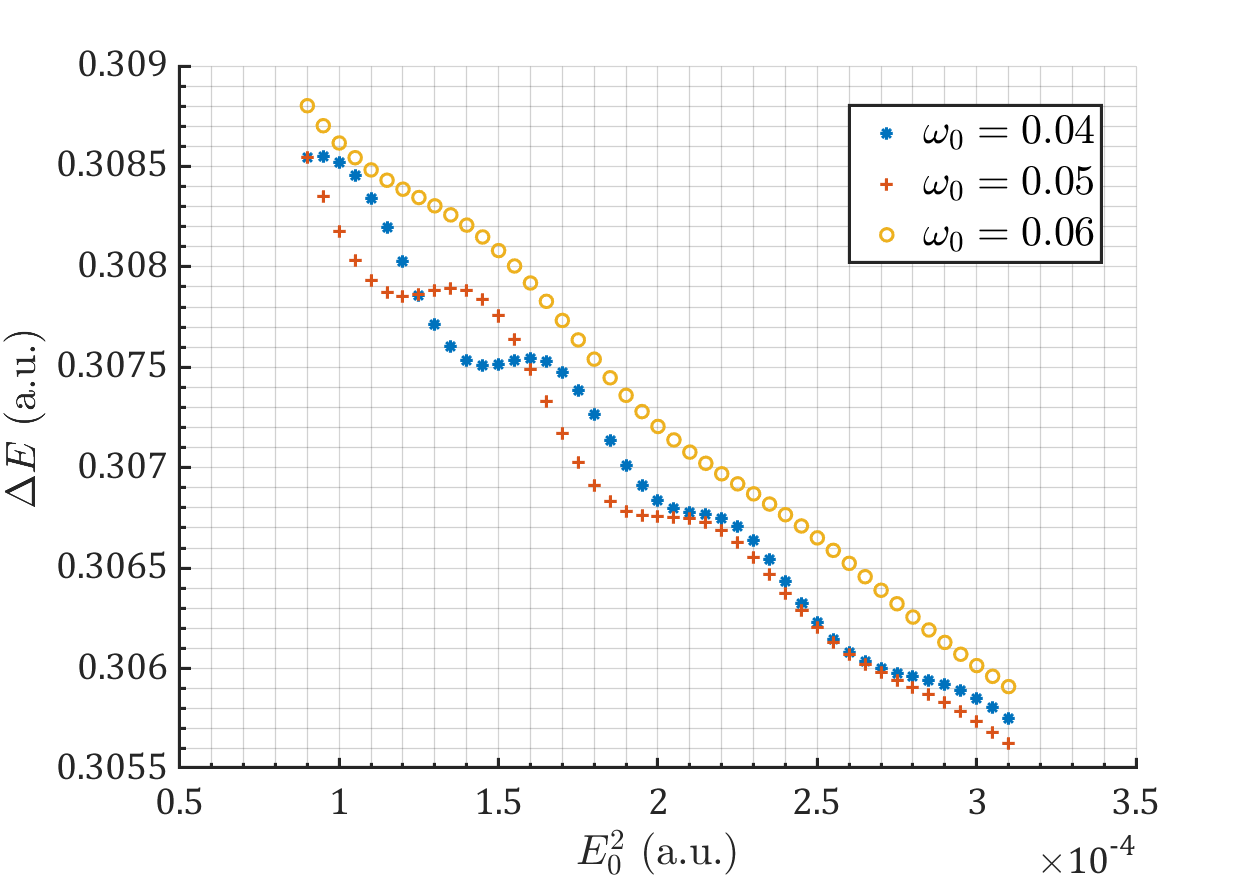}
	\caption{The frequency of the transition from the ground to the first excited state modified by the Stark shift, for different fundamental intensities and frequencies.}
	\label{fig:Stark}
\end{center} 
\end{figure}

\begin{figure*}
\begin{center}
    	\includegraphics[width=0.4\textwidth]{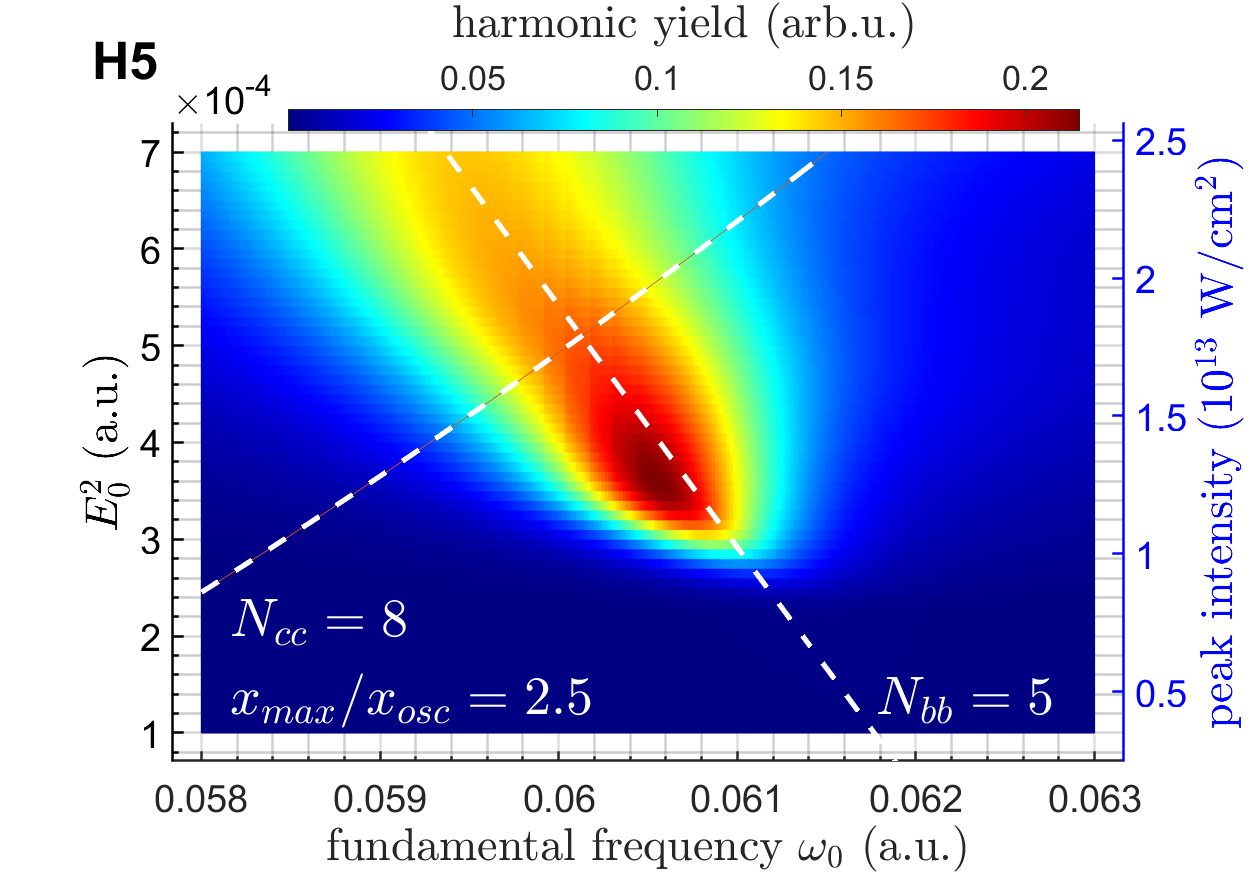}
        \includegraphics[width=0.4\textwidth]{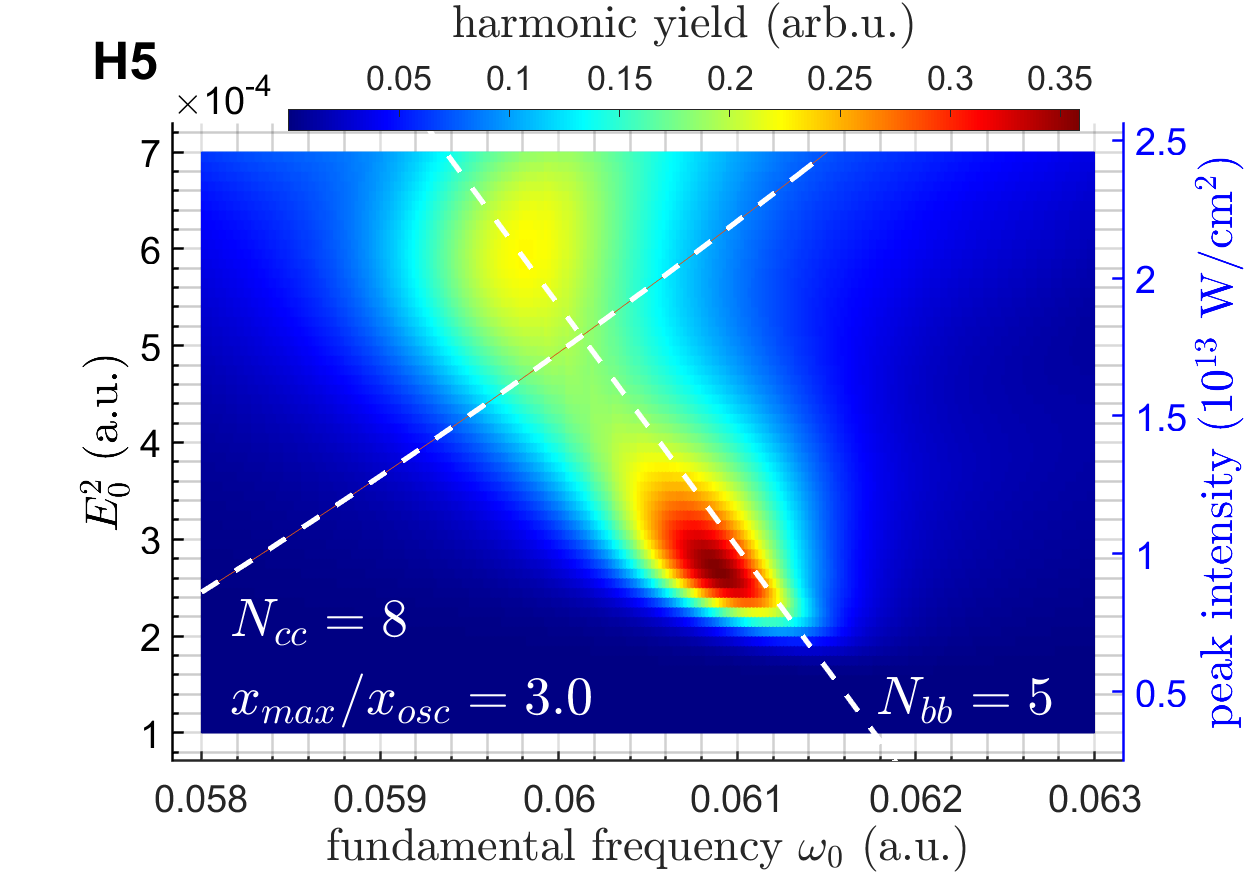}
        \includegraphics[width=0.4\textwidth]{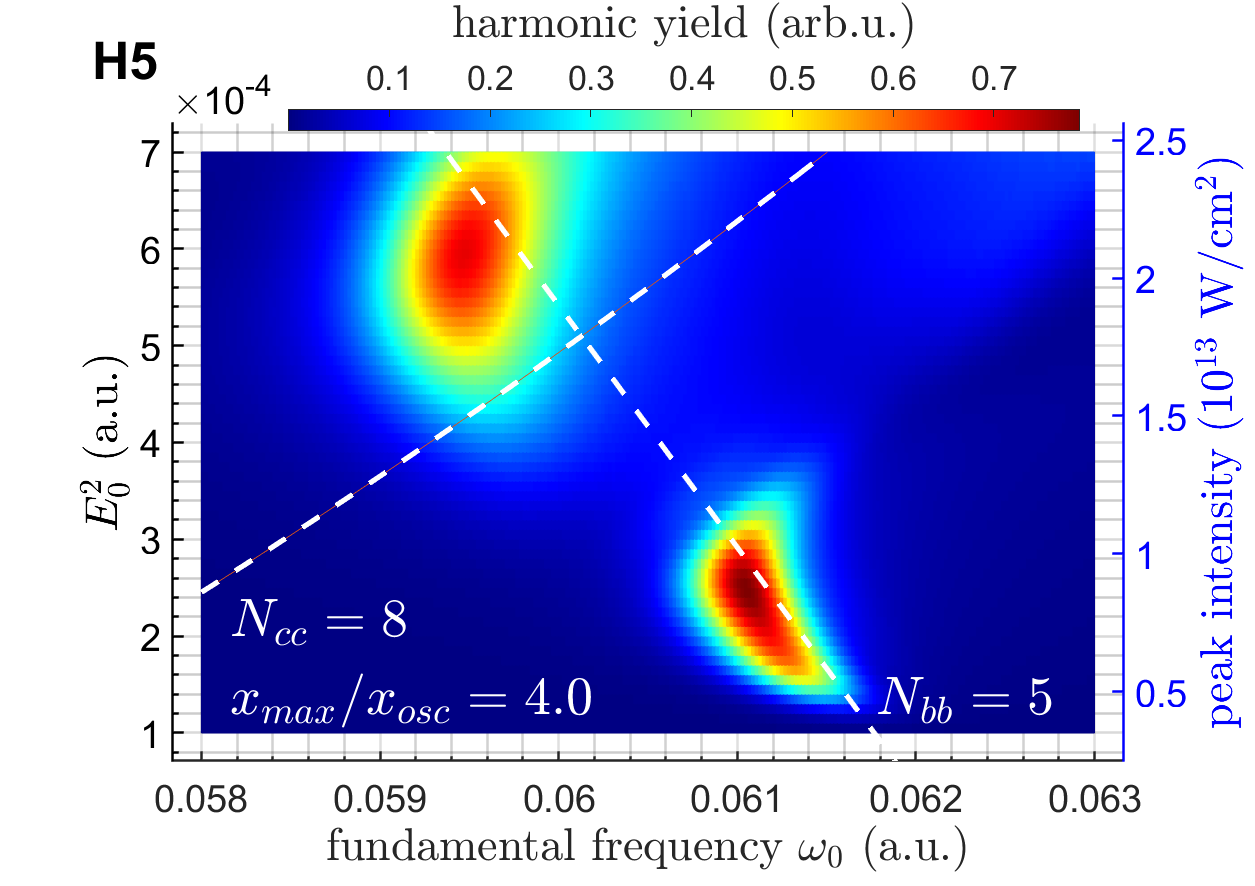}
        \includegraphics[width=0.4\textwidth]{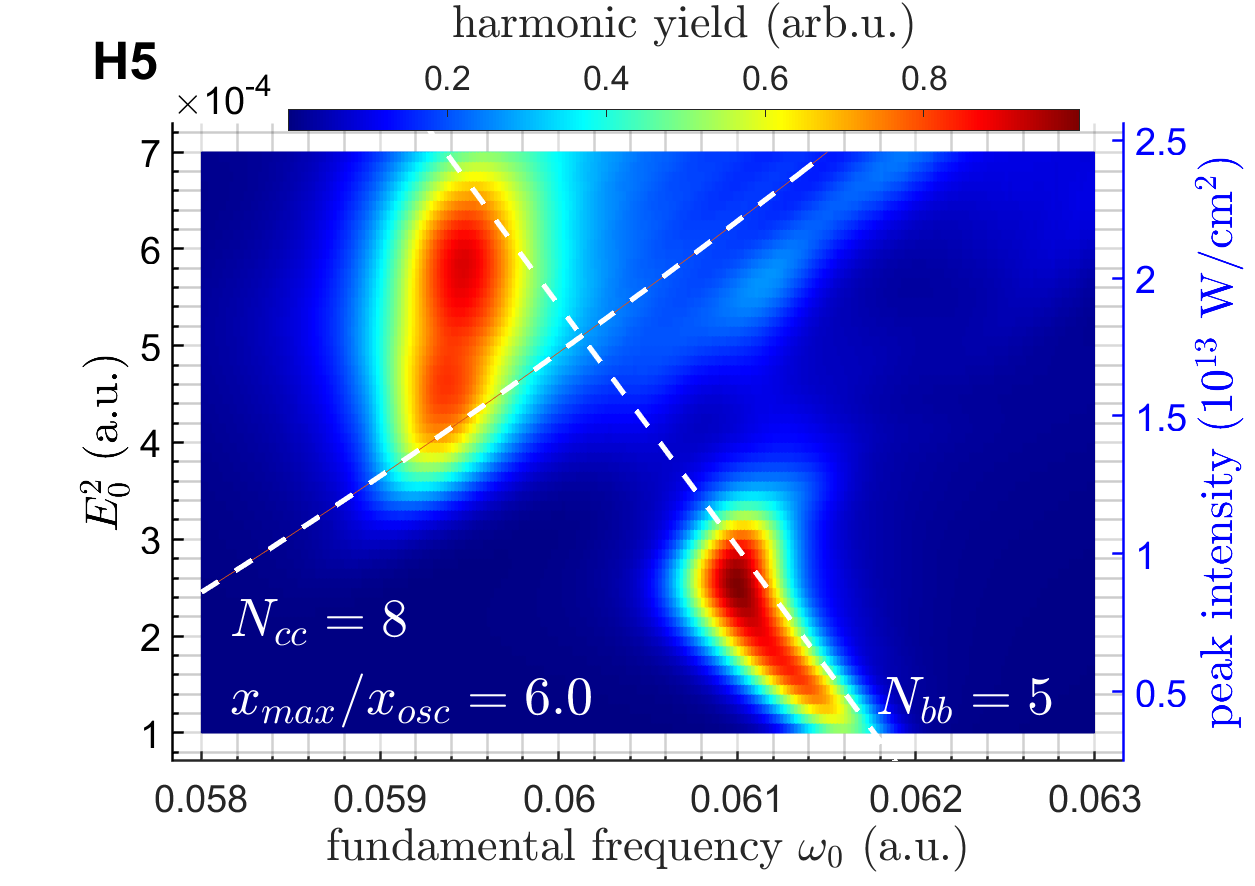}
	\caption{Gradual appearance of the channel-closing effect as the size of the computational box is increased. Numbers of photons for channel-closing $N_{cc}$ and bound-bound transition $N_{bb}$ resonances are presented in the graphs. Size $x_{\max}$ is indicated on each panel in units of the oscillating radius.}
	\label{fig:frustrated_tunneling}
\end{center} 
\end{figure*}

\subsection{Interference of channel closing and bound-bound transition resonance}

In this section, we consider the ionization probability and harmonic yields for lower peak laser intensities, namely ranging from $3.5 \times 10^{12}$ to $2.8 \times 10^{13}$ W/cm$^2$, see Fig.~\ref{fig:harmonics_and_ionization_low_intensity}. A notable feature of this regime is the interference between two types of resonances: the channel-closing resonances discussed above and multiphoton bound-bound transition resonances, corrected by the dynamic Stark effect. The region where the interference occurs is shown in  Fig.~\ref{fig:avoided_crossing_area}.

To verify that the second type of resonances indeed originates from the dynamic Stark effect, we analyze how the relevant bound-state energies shift in the presence of an intense laser pulse. For this purpose, we solve the TDSE \eqref{eq: TDSE} with the initial wavefunction prepared as an equal-amplitude superposition of the field-free ground state and the first excited state:
\begin{equation*}
\Psi(x, 0) = \frac{1}{\sqrt{2}} \bigl( \phi_0(x) + \phi_1(x) \bigr).
\end{equation*}
This choice enhances the visibility of the excited-state contribution and allows for more accurate tracking of its evolution.
From the resulting time-dependent wavefunction $\Psi(x,t)$, we isolate a temporal window near the peak of the laser pulse (specifically, within the FWHM duration), where the field-induced effects are most pronounced. We construct the energy representation by performing a Fourier transform of this wavefunction  with respect to time to obtain $\widetilde{\Psi}(x, E)$  and subsequently integrate its squared modulus over the spatial coordinate to obtain the energy spectrum:
\begin{equation*} 
    S(E) = \int |\widetilde{\Psi}(x, E)|^2 \, dx.
\end{equation*}
The function $S(E)$ exhibits a series of sharp peaks corresponding to the energies of the field-dressed states. To extract these energies with improved precision, we compute the first moment of each peak according to
\begin{equation*} 
    E_i = \frac{\int_{\text{peak}} E \, S(E) \, dE}{\int_{\text{peak}} S(E) \, dE},
\end{equation*}
where the integration is performed over the vicinity of the $i$th peak.
This procedure allows us to quantify the shift of the energy levels as a function of the peak laser intensity. In particular, Fig.~\ref{fig:Stark} shows the energy of the transition from the ground to the first excited state $\Delta E= E_1-E_0$ as a function of the squared amplitude of the field. We see that the transition frequency depends on the field's amplitude due to the Stark shift of the levels. This effect does not depend significantly on the laser frequency, so it is non-resonant. 

    As mentioned earlier, the numerical box size was chosen large enough to accommodate, during its propagation in the continuum, all parts of the electron wave packet responsible for the investigated effects. By deliberately reducing the box size, we can exclude the 'very long' electron's trajectories, thus suppressing the channel-closing effect.
   The results are presented in Fig.~\ref{fig:frustrated_tunneling} for the range of laser field parameter values where both resonances occur: the channel-closing one requires eight laser photons $N_{cc}=8$, while the resonance with the bound-bound transition requires five photons $N_{bb} \equiv \Delta E/ \omega_0=5$ of similar energy. For small computational boxes, we see only the contribution of the resonance with the bound-bound transition. For larger boxes, the picture is affected by an interplay with the channel-closing resonance. 
   
Our study shows that the bound-bound transition resonance can lead to spectral features fundamentally similar to the well-known Fano line shape ~\cite{Fano1961} in the single-photon photoionization spectrum near the autoionizing state (AIS). The non-resonant HG due to recombination of the electron wave packets traveling along 'long' trajectories acts as a continuum contribution in the Fano picture. The excited resonant state acts as the AIS. It should be noted that the excited state is unstable but decays due to photoionization. This broadens its spectral line, making it similar to the AIS case.

\begin{figure}
\begin{center}
    	\includegraphics[width=0.4\textwidth]{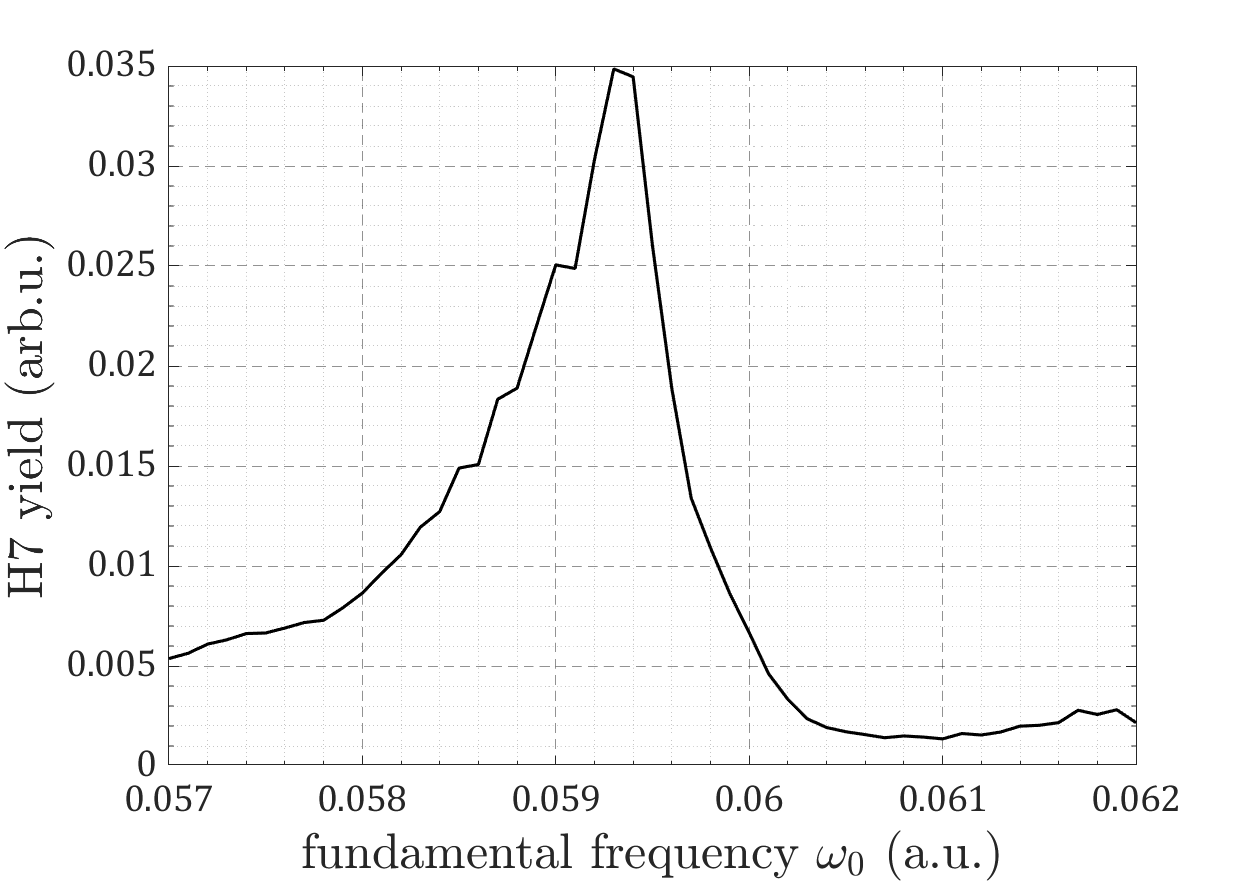}

	\caption{ H7 signal as a function of the fundamental frequency near the 5-photon resonance with the bound-bound transition (cut of Fig.~\ref{fig:avoided_crossing_area}b) for the $E_0^2 = 5 \times 10^{-4}$~a.u.}
	\label{fig:cut}
\end{center} 
\end{figure}

As an example, Fig.~\ref{fig:cut} presents the H7 signal as a function of the fundamental frequency near $N_{bb}=5$. One can see a very typical asymmetric line shape.  It should be noted that Fano-type resonances in above-threshold HHG in the vicinity of AIS are actively studied~\cite{Ganeev2006, Strelkov_2025, Shiner2011, Singh2026}.

The interference of numerous long quantum trajectories leads to channel-closing features in the below-threshold HG spectrum. The shape of the Fano-type resonance line in the presence of such 'continuum' contribution can be rather complex. In Figs.~\ref{fig:harmonics_and_ionization_low_intensity}(b) and ~\ref{fig:avoided_crossing_area}(b) near $\omega_0 = 0.06$ a.u. ($N_{bb}=5$) we see that when the channel-closing maximum falls into the Fano dip, the former naturally disappears. This leads to the 'avoided crossing' picture in Figs.~\ref{fig:harmonics_and_ionization_low_intensity}(b) and ~\ref{fig:avoided_crossing_area}(b) near $\omega_0 = 0.06$ a.u. However, near  $\omega_0 = 0.044$ a.u. ($N_{bb}=7$) the 'avoided crossing' of the resonances looks different. This can be attributed to another shape of the Fano-type line for this resonance.

\section*{Conclusion}
In this paper, we simulate photoionization and HG in an intense femtosecond laser pulse by numerically solving the TDSE for a model xenon atom in a  field. Calculations are performed for variuos  laser pulse frequencies (photon energies from $1.1$ eV to $1.9$ eV) and intensities (up to $1.4 \times  10^{14}$ W/cm$^2$). These parameter ranges cover multiphoton, intermediate, and tunneling ionization regimes. We show that both the photoionization and HG yields are enhanced by two systems of resonances: the channel-closing resonances and the bound-bound transition resonances affected by the Stark-shift of the states.
We find a remarkable behavior of the yields in the range of the laser parameter values where both resonant conditions are satisfied (for different numbers of photons): in this region, the efficiency is enhanced, but in the meantime there is a pronounced dip near the exact resonance intersection. This yields a pattern similar to the ’avoided crossing’ phenomenon well-known in molecular physics and nonlinear dynamics. In contrast to a molecule, where both resonance systems are always present, in our case the ’contribution’ of one of the systems can be tuned. We do this numerically by controlling the size of the computational box and hypothesize that this can be accomplished experimentally by observing the harmonic signal at different angles to the laser beam axis. Furthermore, we show that HG at certain fundamental frequencies and intensities is enhanced due to both resonance mechanisms, leading to a significant increase in the generation efficiency. This is especially important in the case of the 7-photon resonance of the 1030 nm (0.044 a.u., 1.2 eV) radiation of the ytterbium laser because the resonant 7th harmonic generation in xenon is a promising way to efficiently produce the 148 nm UV radiation practically important for operation of Thorium-229 nuclear clock.

\section*{Data availability statement}
The data that support the findings of this article are openly available~\cite{data_Krupin}.

\section*{Acknowledgment}
This work was supported by the Russian Science Foundation (Project No. 25-72-10172).

\bibliography{lit} 

\end{document}